\documentclass[12pt]{article}
\begin{document}
\newcommand{\br}{{\bf r}}
\newcommand{\grad}{\mbox{\boldmath$\nabla$}}
\newcommand{\bdiv}{\mbox{\boldmath$\nabla\cdot$}}
\newcommand{\curl}{\mbox{\boldmath$\nabla\times$}}
\newcommand{\bcdot}{\mbox{\boldmath$\cdot$}}
\newcommand{\btimes}{\mbox{\boldmath$\times$}}
\newcommand{\btau}{\mbox{\boldmath$\tau$}}
\newcommand{\btheta}{\mbox{\boldmath$\theta$}}
\newcommand{\bmu}{\mbox{\boldmath$\mu$}}
\newcommand{\bepsilon}{\mbox{\boldmath$\epsilon$}}
\newcommand{\bcj}{\mbox{\boldmath$\cal J$}}
\newcommand{\bcf}{\mbox{\boldmath$\cal F$}}
\newcommand{\bbeta}{\mbox{\boldmath$\beta$}}
\newcommand{\lbar}{\lambda\hspace{-.09in}^-}
\newcommand{\bcp}{\mbox{\boldmath$\cal P$}}
\newcommand{\bco}{\mbox{\boldmath$\omega$}}
\newcommand{\brho}{\mbox{\boldmath$\rho$}}
\newcommand{\balpha}{\mbox{\boldmath$\alpha$}}
 \title{Comment on “Magnetic moments in the
Poynting theorem, Maxwell equations,
Dirac equation, and QED''}
  \author{Jerrold Franklin\footnote{Internet address: Jerry.F@TEMPLE.EDU}\\
Department of Physics\\ Temple University, Philadelphia, PA 19122}
  \date{\today}
   \maketitle
\begin{abstract}
The paper, ``Magnetic moments in the
Poynting theorem, Maxwell equations,
Dirac equation, and QED'', adds a new magnetic interaction energy,
 from a  `dual magnetic monopole model', 
 to Poynting's theorem to add magnetic moment interaction energy that, presumably,
was absent in Poynting's theorem. We show in this Comment that this leads to errors that invalidate the paper.
 \end{abstract}

The paper, “Magnetic moments in the
Poynting theorem, Maxwell equations,
Dirac equation, and QED'', starts with the statement,``The conventional form of the [Poynting] theorem\footnote{The Poynting theorem is discussed in most advanced electromagnetism books. 
 See, for instance \cite{jdj}.}
 does not take into account
the interaction of the magnetic moment of a particle,
such as an electron, with an inhomogeneous magnetic field.''

How can the paper say that when its own statement of Poynting's theorem in its Eqs.~(4) and (5) includes the energy density,
\begin{eqnarray}
u&=&\frac{1}{\epsilon_0}\left({\bf E\bcdot E+B\bcdot B}\right),
\end{eqnarray}
that includes  the interaction of two magnetic fields, one of which could be the field
of `the magnetic moment of a particle', and the other, `an inhomogeneous magnetic field'? 
Is the paper saying that the magnetic field of a magnetic dipole is not included in the magnetic field in Eq,~(1), and the paper's equation (5)? (What a strange sentence. A magnetic field that is not included in the magnetic field.)

The paper then states that, ``Magnetic moment sources in the Maxwell equations are....
based on the current loop model of the magnetic dipole.''
But the magnetic field in Eq.~(1) above also includes other magnetic dipole sources. Any electric current, not just a current loop, produces a magnetic field, for which a multipole expansion can include a magnetic dipole moment contribution.
Also, any intrinsic magnetic moment\footnote{See, for instance, Section 7.11.5 of \cite{jf}.} would have a magnetic field. 
These magnetic interactions are required by the laws of electromagnetism, and should be included in the magnetic field in Poynting's theorem.  
We will refer to these magnetic fields as `the traditional magnetic field of a magnetic dipole'.

The paper's apparent denial of magnetic interactions in its energy density, $u$, leads it to pose the question, ``How could the Poynting theorem and thus the Maxwell
equations omit a magnetic moment source term that
should be included for 160 years?''

The answer is that those magnetic interaction terms {\it had not} been omitted, except by the paper.
How could the magnetic field of a magnetic dipole been left out of the magnetic field?
The paper claims that these important terms had been carelessly left out.  But, if that were the case, the many calculations based on Poynting's theorem in those 160 years would have disagreed with experiment. 

Because of its mistaken conclusion that magnetic interactions had not already been included in Poynting's theorem, the paper added a term, $-{\bf K}\bcdot c{\bf B}$
to the right hand side of Poyntings theorem. 
The vector,
$\bf K$, is called a `magnetic current density', for the dual magnetic monopole model, and is given by
Eq.~(25) of the paper,
\begin{eqnarray}
 {\bf K(x)}&=&-({\bf v}/c) {\bf m}\bcdot\grad\delta({\bf x - x_0}).
 \end {eqnarray}

For some reason, to correct the assumed absence of traditional magnetic dipole terms, the paper adds, not the claimed missing terms, but a presumed energy density of its 
`dual magnetic monopole model', in which a magnetic dipole is assumed to be like an electric dipole, but uses North and South magnetic monopoles instead of plus and minus electric charges.
Such a model has no experimental evidence, and no theoretical support in textbooks.
The only place I have seen it is in this paper.

 The paper has added its `dual magnetic monopole model' energy density rather than the traditional magnetic moment energy density it had said was missing. That means that the magnetic moment energy the paper had claimed was absent {\it would still be absent.}
 %The stated purpose of the paper to remedy this absence seems to have been forgotten.
 Why  didn't the paper restore the traditional dipole energy it had said was missing? 
 
 Instead it added its `dual magnetic monopole model' energy.
 Adding the dual magnetic monopole interaction introduces new problems in an attempt to cure a problem that wasn't there.

 The paper has made the same mistake it made at the beginning. 
 By adding the energy density of its dual magnetic moment model to the right hand side of Poynting's theorem, is the paper saying that the magnetic field of its model is not part of the magnetic field,
  $\bf B$, on the left hand side?  Again, how can a magnetic field be kept out of the magntic field?

 Adding the energy density of any magnetic dipole to the right hand side, instead of including 
 it in $\bf B$ on the left hand side of Poynting's theorem
 is wrong. This means, the subsequent equations that  use that model are wrong, 
 and the paper is wrong.
    
Actually, it is good news that so much in the paper is wrong because
that means Maxwell’s equations and Poynting’s theorem are still correct after over a
hundrend years of usage.

\end{document}